\documentclass[%
reprint,
superscriptaddress,
nobibnotes,
amsmath,amssymb,
aps,
prd,
reprint,
]{revtex4-1}

\usepackage{graphicx}%
\usepackage{dcolumn}%
\usepackage{bm}%
\usepackage{multirow}
\usepackage{booktabs}
\usepackage[normalem]{ulem}
\usepackage{xcolor}

\newcommand{\Splus}[0]{\hat{S}_{+}}
\newcommand{\Ssquare}[0]{\hat{S}^{2}}
\newcommand{\ket}[1]{|{#1}\rangle}
\newcommand{\bra}[1]{\langle{#1}|}
\newcommand{\braket}[2]{\langle{#1}|{#2}\rangle}
\newcommand{\bdX}[0]{\boldsymbol{X}}

\newcommand{\penaltya}[0]{\overline{P}_{\Splus}}
\newcommand{\penaltyb}[0]{P_{\Splus}}

\begin{document}

\title{Symmetry enforced solution of the many-body Schrödinger equation\\ with deep neural network}

\author{Zhe Li}
\thanks{These authors contribute equally.}
\email{lizhe.qc@bytedance.com}
\affiliation{ByteDance Research}

\author{Zixiang Lu}
\thanks{These authors contribute equally.}
\affiliation{ByteDance Research}
\affiliation{Peking University}

\author{Ruichen Li}
\thanks{These authors contribute equally.}
\affiliation{ByteDance Research}
\affiliation{Peking University}

\author{Xuelan Wen}
\affiliation{ByteDance Research}

\author{Xiang Li}
\affiliation{ByteDance Research}

\author{Liwei Wang}
\thanks{wanglw@cis.pku.edu.cn}
\affiliation{Peking University}

\author{Ji Chen}
\thanks{ji.chen@pku.edu.cn}
\affiliation{Peking University}

\author{Weiluo Ren}
\thanks{renweiluo@bytedance.com}
\affiliation{ByteDance Research}

\begin{abstract}
The integration of deep neural networks with the Variational Monte Carlo (VMC) method has marked a significant advancement in solving the Schrödinger equation.
In this work, we enforce spin symmetry in the neural network-based VMC calculation with modified optimization target.
Our method is designed to solve for the ground state and multiple excited states with target spin symmetry at a low computational cost.
It predicts accurate energies while maintaining the correct symmetry in strongly correlated systems, even in cases where different spin states are nearly degenerate.
Our approach also excels at spin-gap calculations, including the singlet-triplet gap in biradical systems, which is of high interest in photochemistry.
Overall, this work establishes a robust framework for efficiently calculating various quantum states with specific spin symmetry in correlated systems, paving the way for novel discoveries in quantum science.

\end{abstract}

\maketitle
Accurately characterizing subtle electron-electron correlation is essential for understanding multi-reference systems in quantum physics and chemistry, especially in areas such as catalysis, photochemistry, and superconductivity.
On one hand, widely used Kohn-Sham density functional theory (KS-DFT) \cite{khon-sham-dft} falls short in accurately accounting for static correlation in multi-reference systems~\cite{dft_limitation}.
This leads to a phenomenon known as the symmetry dilemma \cite{symmetry-dilemma-1, symmetry-dilemma-2}, where the spin symmetry broken solution, an unphysical state, achieves lower energy result.
On the other hand, there are a number of wavefunction methods that are good at capturing static correlation \cite{review_mrci, review-excited-mr-method}.
However, these methods come with their own challenges and limitations, such as their prohibitive computational scaling and the requirement of expertise knowledge to select appropriate active spaces.

Neural network-based variational Monte Carlo (NNVMC) method offers an attractive alternative~\cite{hermann2023ab}. 
Permutation-equvariant neural networks with a large number of parameters have been employed to describe the wavefunction, which can be optimized based on the variational principle.
Its efficacy in solving the ground-state wavefunction has been demonstrated extensively, showing a promising route to surpass the gold standard quantum chemistry method in terms of accuracy while maintaining reasonable computational costs \cite{carleo-nn1, deepwf, paulinet, ferminet, deepsolid,  diluo-backflow, gerard2022gold, gao-pesnet, gao-moon, psiformer, lin2023explicitly, lapnet}.
Even for challenging multi-reference systems, NNVMC exhibits very accurate results compared with other state-of-the-art methods \cite{fu2024variance, penalty_paulinet, nqmc}.
However, optimizations in NNVMC are not guaranteed to provide quantum states with correct spin symmetry, which is one of the fundamental intrinsic symmetries in nonrelativistic systems.
Specifically, the optimized neural network ansatz in general is not the eigenfunction of the spin square operator $\hat{S}^2$, resulting in what is known as spin contamination~\cite{spin-contamination-1, spin-contamination-2}.
This is particularly common in cases where different spin states are nearly degenerate due to strong electron correlation.

In this work, we expand the scope of NNVMC research by obtaining correct ground and excited states with target spin symmetry.
We introduce a low-scaling penalty term in VMC loss function that enforces spin symmetry with minimal computational overhead. 
\begin{figure*}[t]
    \centering
    \includegraphics[width=\linewidth]{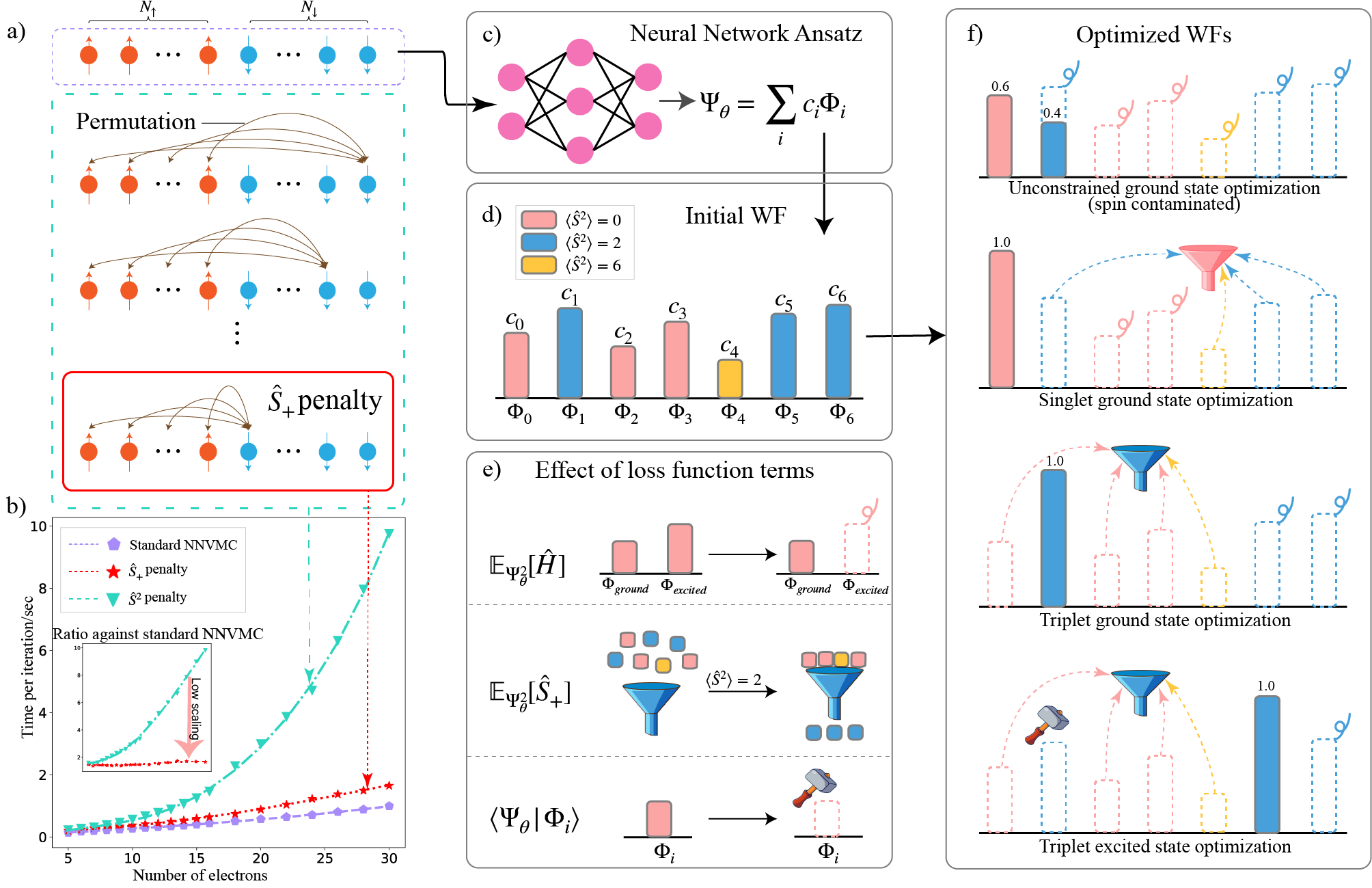}
    \caption{
    \textbf{Framework overview.}
    (a) Comparison of penalty methods to enforce symmetry:
    The cyan frame shows the $\hat{S}^2$ penalty method, requiring evaluating $N_\uparrow N_\downarrow$ data points constructed from swapping every spin-up electron with each spin-down electron in the original input.
    The red frame shows our proposed low-scaling $\hat{S}_{+}$ penalty method, where only a single spin-down electron is interchanged with electrons of the opposite spin, resulting in the evaluation of only $N_\uparrow$ data points.
    (b) Computational runtime comparison on three approaches, namely standard NNVMC, our $\hat{S}_{+}$ penalty and the $\hat{S}^2$ penalty, across systems with different electrons.
    The inset shows the runtime ratio to the standard NNVMC, demonstrating the one-order reduction of computational complexity of $\hat{S}_{+}$ penalty over $\hat{S}^2$ penalty.
    (c) We use neural networks to represent the wavefunction ansatz in real space. 
    (d) The expansion of the neural network wavefunction over the Hamiltonian eigenstates, with bar height indicating amplitude of each eigenstate and bar color indicating different spin symmetries.
    (e) Effect of different components of the loss function in our modified NNVMC process. 
    At the top is the energy value, eliminating state components with higher energy as in standard NNVMC.
    In the middle, we have our proposed $\hat{S}_+$ penalty, retaining only the states with target spin symmetry.
    At the bottom is the overlap penalty~\cite{penalty_wagner1, penalty_wagner2, penalty_paulinet}, eliminating the component corresponding to the specified state.
    (f) The demonstration of NNVMC calculation with enforced spin symmetry.
    The top row illustrates that unconstrained NNVMC may results in spin contamination for ground state calculations.
    The second row shows that $\hat{S}_+$ penalty helps achieve better ground state estimate without spin contamination.
    The last two rows show how our approach can efficiently obtain quantum states with specified spin symmetry.
    }
    \label{fig:sketch map}
\end{figure*}
By incorporating symmetry, we enhance the reliability of optimization processes and achieve improved energy results for the ground state of multi-reference systems.
This is demonstrated with ethylene in its perpendicular geometry.
Our approach also achieves higher accuracy and efficiency for excited state calculations. 
It effectively captures all the expected states in the selected atomic spectrum.
Additionally, we achieve chemical accuracy in calculations of the spin gap between the lowest singlet and triplet states in biradical systems, consistently matching experimental results for various molecules.
In summary, our symmetry-enforced method accurately and efficiently describes both ground and excited states, facilitating better understanding of challenging strongly correlated systems and advancing applications across various fields in quantum science.

\section{Results}
\subsection{Spin symmetry enforcement}\label{subsec:Splus}
In this section, we describe how to enforce spin symmetry in NNVMC, i.e. enforce the wavefunction ansatz to be an eigenstate of the spin square operator $\Ssquare$ with eigenvalue $s(s+1)$.
This can improve the reliability of the optimization process and yield better results.
To begin with, we constrain our ansatz to only represent quantum states with spin value lower bounded by the target value $s$, by assigning the spin magnetic quantum number $s_z = s$.
This is feasible for most NNVMC ansatz by adjusting the number of spin-up and spin-down electrons.
To further enforce spin at the target value $s$, we introduce a new penalty term into the loss function for the NNVMC optimization process. 
If we don't take computation complexity into account, then it is straightforward to design a penalty term involving $\hat{S}^2$, which can be calculated with its local value \cite{spin-square-0, spin-square-1, spin-square-2, spin-square-3}:

\begin{equation}\label{eq:local-spin-square}
    S^2_L(\boldsymbol{X}) = \frac{N}{2} + \frac{(N_{\uparrow} - N_{\downarrow})^2}{4} - \sum_{\alpha, \beta} \frac{\hat{\mathbb{P}}_{[\alpha, \beta]} [\Psi](\boldsymbol{X})}{\Psi(\boldsymbol{X})},
\end{equation}
where $\Psi$ is the neural network wavefunction and the variable $\boldsymbol{X}$ is the collection of three dimensional coordinates of $N$ electrons, $\boldsymbol{X} \equiv [\boldsymbol{x}_1, \cdots, \boldsymbol{x}_N]$.
The permutation operator $\hat{\mathbb{P}}_{[\alpha, \beta]}$ swaps the coordinates of one spin-up electron with another spin-down electron indicated by $\alpha$ and $\beta$, respectively, namely changing the input of $\Psi$ from $[\boldsymbol{x}_1, \cdots, \boldsymbol{x}_{\alpha}, \cdots, \boldsymbol{x}_{\beta}, \cdots, \boldsymbol{x}_N]$ to $[\boldsymbol{x}_1, \cdots, \boldsymbol{x}_{\beta}, \cdots, \boldsymbol{x}_{\alpha}, \cdots, \boldsymbol{x}_N]$.
We denote $N_\uparrow$ and $N_\downarrow$ as the numbers of spin-up and spin-down electrons respectively.

The calculation of $S^2_L(\boldsymbol{X})$ in Equation \eqref{eq:local-spin-square} is unfortunately very expensive due to $N_\uparrow N_\downarrow$ evaluations of the neural network wavefunction $\Psi$ corresponding to different swaps for electron pairs with opposite spins, which is of order $\mathcal{O}(N^2)$ when $N_\uparrow$ and $N_\downarrow$ are comparable.
In other words, compared to the computational bottleneck of VMC, namely Laplacian operator \cite{lapnet}, the evaluation of $\hat{S}^2$ is even more costly by $\mathcal{O}(N)$.
Therefore, if we directly add $\hat{S}^2$ in the loss function, for instance 
$\mathbb{E}[(S^2_L(\boldsymbol{X}) - s(s+1))^2]$ as suggested by \citet{spin-penalty-1}, then the computational complexity of the whole VMC process will be increased with $\mathcal{O}(N)$.
We refer to this penalty as $\hat{S}^2$ penalty and  illustrate it in figure \ref{fig:sketch map}(a) with the cyan dashed frame, which shows all the required inputs to calculate the spin square value.

In order to achieve better efficiency, we design a new penalty term, called $\hat{S}_+$ penalty, to eliminate the extra $\mathcal{O}(N)$ computational overhead while still able to enforce the spin symmetry effectively. 
The $\hat{S}_{+}$ penalty takes effects through the property of the raising operator $\hat{S}_{+}$, which annihilates the spin state polarized to the $z$-axis, 
\begin{equation*}
\hat{S}_{+} |S=s, S_z=s\rangle = 0.    
\end{equation*}
This property implies that we can derive the eigenstate of $\Ssquare$ through enforcing $\Splus\ket{\Psi}$ to $0$. 
It can be further shown that $\hat{S}_{+}|\Psi\rangle=0$ is equivalent to the following condition (See more details in Section \ref{subsec:method_spin_penalty}):
\begin{equation}
\label{eq:splus_res}
    \forall \boldsymbol{X}\in\mathbb{R}^{3N}, R_{\beta}(\boldsymbol{X}) \triangleq 1 - \sum_{\alpha} \frac{\hat{\mathbb{P}}_{[\alpha, \beta]} [\Psi](\boldsymbol{X})}{\Psi(\boldsymbol{X})}=0.
\end{equation}
Compared with the calculation of the term  $S^2_L(\boldsymbol{X})$ in equation \eqref{eq:local-spin-square}, our proposed term $R_{\beta}(\boldsymbol{X})$ only involves the first index in the summation and evaluates the wavefunction for $N_\uparrow$ times, thus reducing the additional computational complexity of $\mathcal{O}(N)$.
The improvement of efficiency is confirmed in our calculations empirically, as shown in Figure\ref{fig:sketch map}(b) comparing scaling curves of computational complexity of our $\hat{S}_{+}$ penalty and the $\hat{S}^2$ penalty.
 
Our approach not only suppresses spin contamination in ground state optimization, but also facilitates efficient and accurate calculation of excited states when combined with the existing penalty methods \cite{penalty_wagner1, penalty_wagner2, penalty_paulinet}, as illustrated in Fig~\ref{fig:sketch map}(d-f).
This point will be further demonstrated in the following sections.
\subsection{Improved ground state optimization}
\begin{figure*}[ht]
    \centering
    \includegraphics[width=\linewidth]{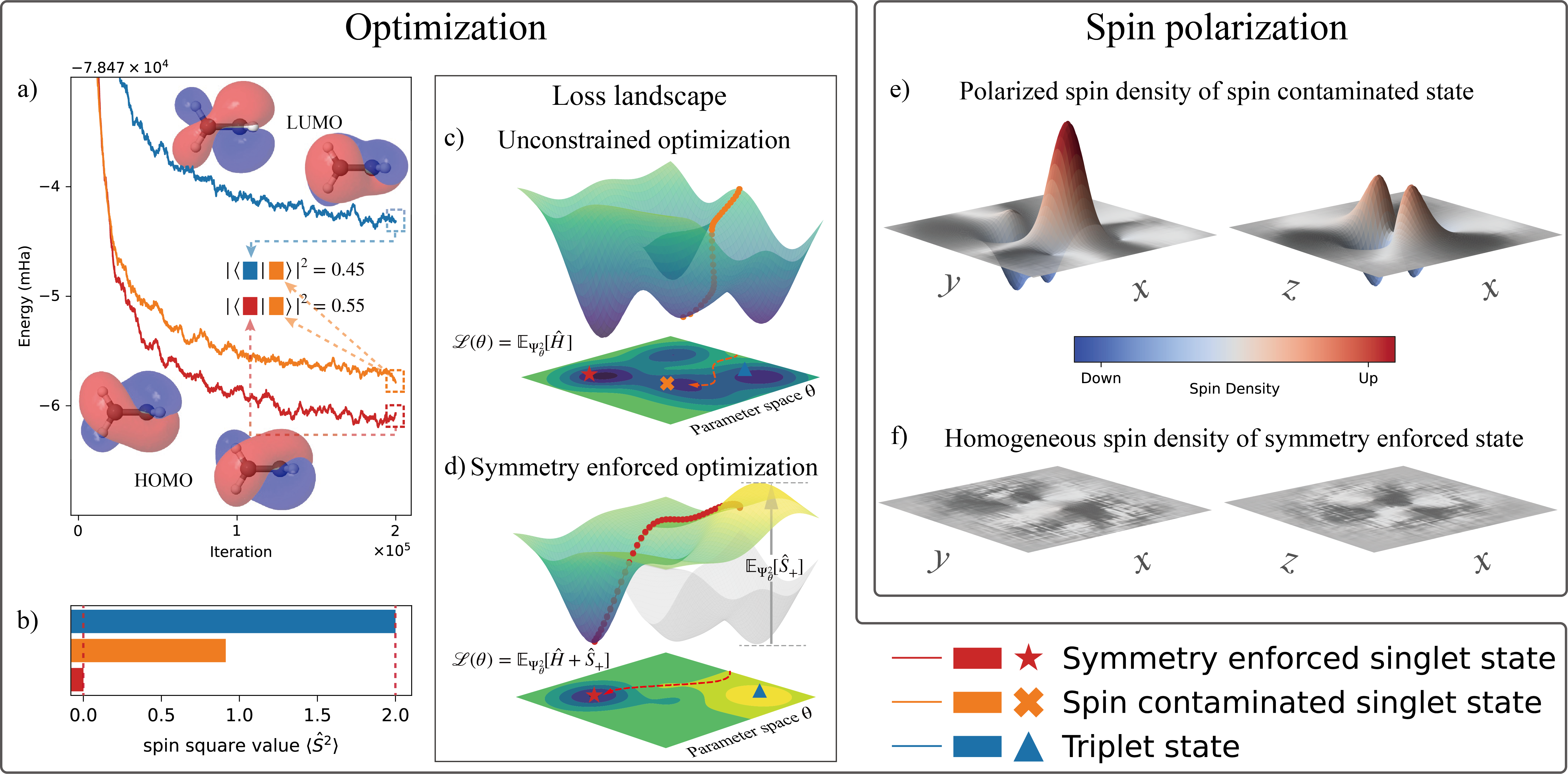}
    \caption{
        \textbf{Enforced symmetry improves ground-state estimate.}
        (a) Training curves of energy.
        The diagram displays the optimization paths for the ground (N) state, a spin singlet state, with red and orange curves corresponding to different optimization methods.
        The orange curve follows unconstrained NNVMC process, whereas the red curve incorporates our $\hat{S}_+$ penalty in the loss function, achieving a lower, namely better, energy level. 
        The blue curve signifies the first excited state, a spin triplet state known as the T state.
        Adjacent to the curves, two singly occupied molecular orbitals of N and T states are displayed, corresponding to the HOMO and LUMO, respectively.  
        The state represented by the orange curve is a superposition of N and T states, with respective square overlap magnitudes of $0.55$ and $0.45$.
        (b) Spin square values.
        Bars represent the spin square value for each state, colored similarly to the curves in panel (a).
        The red and blue bars, with values of $0$ and $2$, indicate states with good spin symmetry, namely singlet and triplet respectively. 
        The orange bar, with a value of $1$, confirms the presence of spin contamination in the state obtained from unconstrained NNVMC process.
        (cd) Loss landscape for neural network training.
        (c) Loss landscape in unconstrained NNVMC optimization, with loss composed solely of an energy term.
        The deepest valley, representing the spin singlet ground state, is marked with a red star on the projection plane, namely the parameter space.
        The first excited state in spin triplet, is situated in the subsequent deepest valley to the right, marked with a blue triangle. 
        The trajectory of the optimization is stuck at one local minima with spin contamination, corresponding to the orange cross.
         (d) The loss landscape is improved when spin symmetry is enforced by integrating $\hat{S}_+$ penalty term into the loss function, which elevates the loss hypersurface near triplet states and eliminates suboptimal local minima.
        This change in landscape leads to better convergence towards the correct state.
        (e)(f) Spin density distribution on the xy- and xz-plane.
        (e) The spin density from unconstrained NNVMC process shows spin polarization near carbon atoms, breaking the homogeneity.
        (f) Enforcing spin symmetry restores a uniform spin density distribution.
    }
    
    \label{fig:ethylene}
\end{figure*}
For multi-reference systems with nearly-degenerate states, NNVMC often gets trapped in local minima, resulting in unsatisfactory quantum states.
The challenge is reflected in a loss landscape cluttered with suboptimal local minima, as illustrated in Fig~\ref{fig:ethylene}(c).
By enforcing spin symmetry, namely incorporating $\hat{S}_+$ penalty into the loss function, the whole loss landscape can be significantly improved, as the states with the wrong spin symmetry are heavily penalized.
Consequently, NNVMC will achieve a smoother and more robust optimization process, reaching target quantum states.

To illustrate this point, we take ethylene as an example, a molecule of paramount importance in scientific research, especially in the fields of photochemistry and quantum dynamics. 
Ethylene is also a renowned challenging system for electronic structure calculations due to the complex interplay of correlation effects and near-degeneracies, especially in its twisted configuration \cite{review_diradical}.
We consider its perpendicular geometry, where the N state (ground state) and the T state (the lowest triplet state) are nearly degenerate \cite{ethylene-conical}.
In our NNVMC calculation, we successfully obtain the correct singlet N state with our $\hat{S}_+$ penalty.
We also obtain the first excited state using the penalty method as in \cite{penalty_wagner1,penalty_wagner2,penalty_paulinet}, which turns out to be a spin triplet, specifically the lowest triplet T state.
Training curves and the spin values are shown in Fig \ref{fig:ethylene} (a-b), with the N state in red and the T state in blue.
Note that the energy gap between N and T states is very small, only 2 mHa, confirming the nearly degenerate nature of this system.
Moreover, in those two states, there are two singly occupied molecular orbitals, namely the highest occupied and lowest unoccupied molecular orbitals (HOMO and LUMO), visualized alongside the corresponding curves in Fig \ref{fig:ethylene}(a).

The open-shell singlet nature of ground state results in an unstable electronic configuration, making it highly chemically reactive.
Due to this reactivity and near-degeneracy, the VMC loss landscape of this system is expected to be highly intricate.
The unconstrained NNVMC calculation, shown as orange curve in Fig \ref{fig:ethylene} (a), optimized to a spin contaminated state with energy between the N and T states, which indicates that the optimization process is indeed trapped at a suboptimal local minimum.
The overlap calculation reveals that this spin-contaminated state is a superposition of N and T states, with respective square magnitudes of $0.45$ and $0.55$.
The corresponding optimization process in the parameter space of loss function is visualized in Fig \ref{fig:ethylene}(c). The optimization process follows the orange arrow and converges to the orange cross, which lies in the middle of the nearly degenerate N and T states. 
On the contrary, as shown in Fig \ref{fig:ethylene}(d), when the $\hat{S}_{+}$ penalty is applied, the loss hypersurface at the T state is substantially lifted and thereby removes several nearby local minima, which explains why we can successfully achieve the global minimum, namely the N state, as mentioned above.

Furthermore, we also investigate the spin density of the obtained states, where we expect homogeneity due to the spin independence of the non-relativisitc Hamiltonian.
Fig \ref{fig:ethylene}(e) shows unphysical spin polarization near the carbon atoms for the state trained with unconstrained NNVMC.
Conversely, when symmetry is enforced, spin distribution of the N state retains spatially homogeneous as expected, as illustrated in Fig \ref{fig:ethylene}(f).
In summary, the symmetry-enforced calculation with $\hat{S}_+$ penalty allows us to achieve target quantum states with better energy and symmetry.

\subsection{Excited states with enforced symmetry}
\begin{figure*}[ht]
    \centering
    \includegraphics[width=\linewidth]{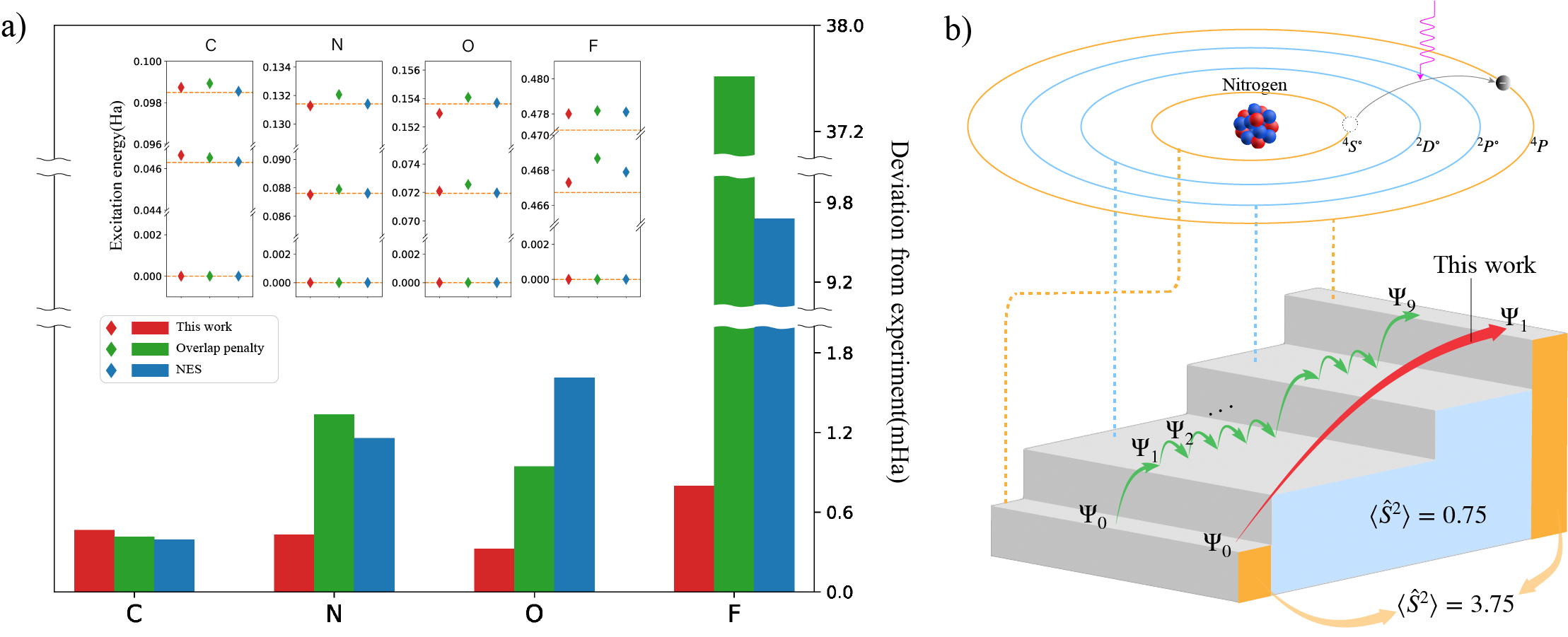}
    \caption{
    \textbf{Symmetry-enforced NNVMC achieves accurate high-lying excited states.}
    (a) 
    We compare atomic spectrum calculated from different methods and demonstrate the advantage of our symmetry-enforced approach.
    We consider the first four energy levels of carbon (C), nitrogen (N), oxygen (O), and fluorine (F), where each energy level may contain multiple degenerate states.
    The bar chart shows the deviation of the fourth excitation energy level from the experimental data \cite{spectrum-handbook}, while the inset shows a comparison of the first three excitation energy levels against the experimental data (yellow reference lines).
    For energy levels with multiple degenerate states, we use the average value across these states.
    Results from this work are colored in red.
    Results for overlap penalty method without symmetry constraints are colored in green.
    Data from \citet{nqmc} are shown in blue color.
    (b) Illustration of the process solving for the $^4P$ state of nitrogen atom. 
    The sequential green arrows represent approaches requiring the computation of all nine lower eigenstates, such as the overlap penalty~\cite{penalty_wagner1, penalty_paulinet, penalty_wagner2} and NES method~\cite{nqmc}.
    In contrast, our symmetry-enforced approach (the red arrow) directly identifies the $^4P$ state as the first quartet eigenstate orthogonal to the ground state, without the need to account for any lower doublet state.}
    \label{fig:atom}
\end{figure*}
\begin{figure*}[ht]
    \centering
    \includegraphics[width=\linewidth]{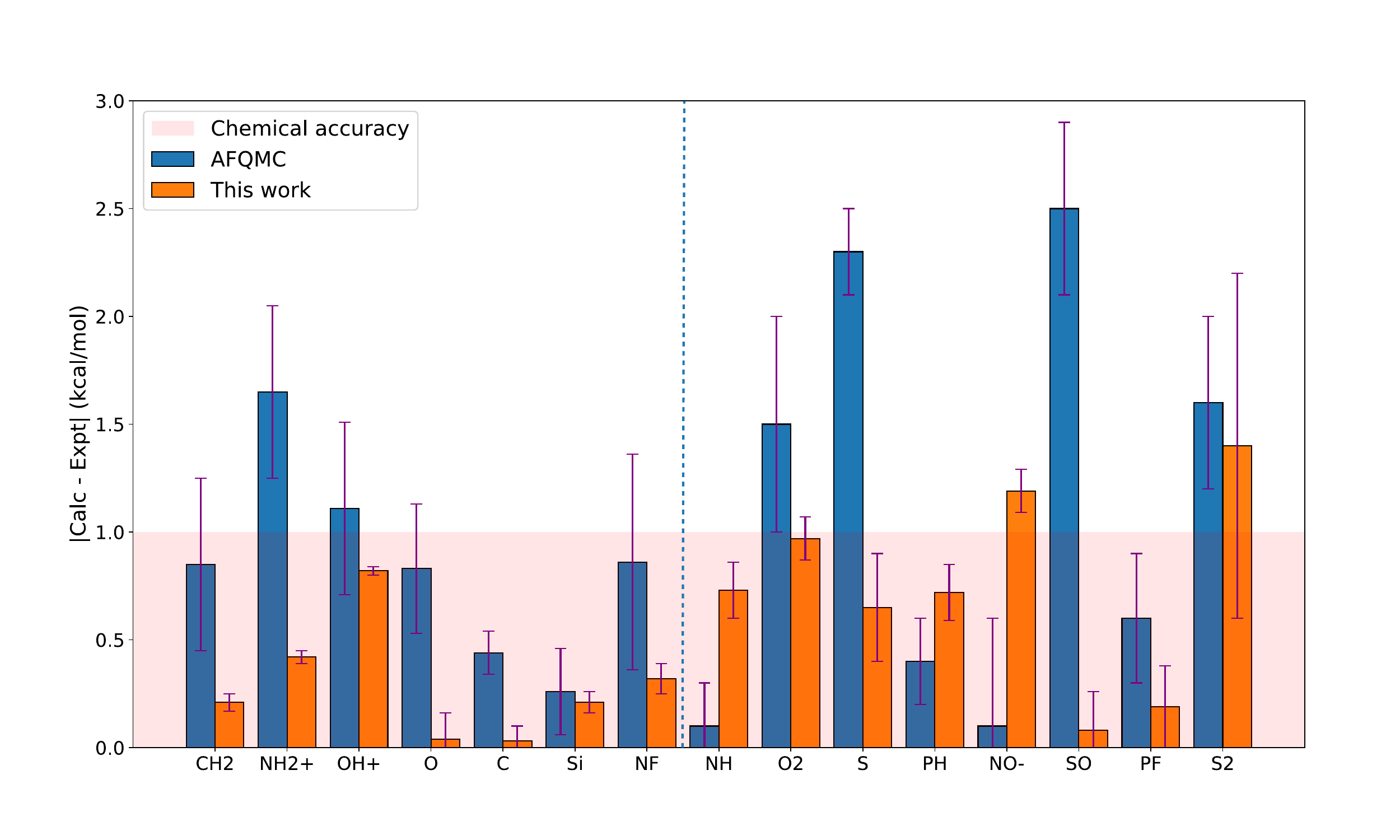}
    \caption{%
    \textbf{S-T gap for a collection of biradical systems.}
    Absolute deviations (in kcal/mol) from experimental references \cite{Biradical_pdft_1_geo, biradical_lee} are displayed.
    The majority of our results are within chemical accuracy.
    Our results are also compared with high-accuracy AFQMC results, \citet{biradical_shee} which uses CASSCF trial(left to the blue dashed line) and \citet{biradical_lee} which uses cRHF trial (right to the blue dashed line).
    }
    \label{fig:biradical}
\end{figure*}

The accurate solution of excited states plays a key role in quantum chemistry, especially in understanding the electronic structures and reactivity of molecules.
In this section, we illustrate how enforcing spin symmetry can significantly improve the accuracy and efficiency of excited state calculations.
Previous studies on excited states in NNVMC framework \cite{penalty_paulinet, nqmc} must account for the full spectrum of intermediate states with energy levels below the target state, causing various optimization issues.
For instance, the overlap penalty method \cite{penalty_paulinet} considers overlap terms for each pair of states involved, with the number of these pairs growing quadratically with the number of excited states.
Despite heavy computational burden, this introduces gradient biases that can misdirect the training process, particularly for high-lying excited states.
The Natural Excited State (NES) method \cite{nqmc} eliminates the need for penalty terms.
However, this advantage comes at the cost of significantly expanding the determinant space as the number of states grows, which still demands considerable computational resources.

In contrast, our symmetry-enforced approach searches for the target states within a far more compact subspace characterized by spin symmetry.
Specifically, we combine the symmetry-enforced method described in Section \ref{subsec:Splus} with the overlap penalty to simulate the high-lying excited states both efficiently and accurately.
This approach, as depicted in Figure \ref{fig:sketch map}(f), narrows the scope of states involved by excluding those with different spins.

We take the nitrogen atom to exemplify our approach as shown in Figure \ref{fig:atom}(b).
Here we consider the lowest $10$ states, which are composed of eight spin doublet states ($\langle \hat{S}^2 \rangle=0.75$) and two quartet states ($\langle \hat{S}^2 \rangle = 3.75$).
The two quartet states, namely ground state and ninth excited state, are respectively denoted as $^4S^{\circ}$ and $^4P$. They are marked as orange steps in Figure \ref{fig:atom}(b).
The remaining eight states reside in the second and third energy levels and are marked as blue steps in Figure \ref{fig:atom}(b). The second energy level is composed by five degenerate $^2D^{\circ}$ states; the third energy level is comprised by three degenerate $^2P^{\circ}$ states.
Our symmetry-enforced approach can directly solve for the $^4P$ state only involving the ground quartet state $^4S^{\circ}$ and skipping all intermediate doublet states.
This results in an order of magnitude reduction in the number of training states and overlap penalty terms, from 10 to 2 and from 20 to 1 respectively, which enhances significantly both the efficiency and accuracy.

As illustrated in Figure \ref{fig:atom}(a), our approach delivers superior results in computing the lowest ten states of carbon, nitrogen, oxygen, and fluorine atoms, outperforming both the sole overlap penalty method and NES.
While all three methods reach chemical accuracy for the first three low-lying energy levels, our method shows clear advantages in computing high-lying excited states.
For instance, the other methods exhibit significant deviations at the fourth energy level for nitrogen and oxygen atoms, and even converge to incorrect states for the fluorine atom. 
Notably, our method maintains chemical accuracy across all tested energy levels, proving its capability and reliability in treating a wide range of excited spin states.

\subsection{Biradical systems} 

In this section, we demonstrate another application of our approach on biradical systems, which are of great scientific interest due to their potential in next-generation organic photovoltaics and molecular magnets~\cite{biradical_photo, sing_fission, review_diradical}.
The singlet-triplet (S-T) gap in biradicals is a key quantity that governs their reactivity, photophysical behavior, and magnetic properties.
Nonetheless, accurately determining the S-T gap in biradicals has historically posed a significant challenge within quantum chemistry due to their complex multireference character \cite{Biradical_casscf, Biradical_cc, Biradical_pdft_1_geo, Biradical_pdft_2, Biradical_RPA1, Biradical_RPA2, Biradical_RPA3, Biradical_acse}.

Here we investigate the S-T gap with $\hat{S}_+$ penalty over a wide range of biradical systems.
Those systems have been recently studied in \cite{biradical_lee, biradical_shee}, where spin-projected Auxiliary-Field Quantum Monte Carlo (AFQMC) is applied to precise S-T gap calculations.
As illustrated in Figure \ref{fig:biradical}, it is evident that our results exhibit accuracy that is comparable to, and often exceeding, that of AFQMC. 
Notably, the majority of our results are within chemical accuracy, \textit{i.e.}, 1 kcal/mol accuracy. 
 Furthermore, an additional advantage of our method is its independence from prior chemical knowledge, such as how to  select active space or basis set.
Consequently, these findings underscore the robustness and accuracy of our method for accurately determining the S-T gap in biradical systems.

\section{Discussion}
Our research presents a significant advancement in NNVMC by incorporating spin symmetry into the optimization process.
We propose a novel $\hat{S}_{+}$ penalty, which allows efficient and accurate calculation of target quantum states with specified spin symmetry.
In addition to the superior efficiency of $\hat{S}_{+}$ penalty, our approach has another advantage of effectively integrating with the overlap penalty \cite{penalty_wagner1, penalty_paulinet, penalty_wagner2} when targeting excited states.
This combination avoids calculations of intermediate states with unnecessary spin symmetry, thereby enhancing the efficiency and accuracy of excited-state calculations.
It is also noteworthy that our method is independent of the choice of wavefunction ansatz, making it applicable to conventional ansatz as well. 
Additionally, by enforcing spin symmetry and orthogonality, NNVMC can generate quantum states without components of undesirable states, making it an good starting point for subsequent Diffusion Monte Carlo (DMC) calculations \cite{nndmc1, nndmc2}.

Concurrent research by \citet{spin-penalty-3} also explores spin-related penalty in NNVMC, similar to the traditional $\hat{S}^2$ penalty \cite{spin-penalty-1}.
Our $\hat{S}_{+}$ penalty significantly reduces the computational complexity compared to those penalties that directly depend on $\hat{S}^2$, which allows calculations for larger and more practical systems.
Furthermore, compared to NES~\cite{nqmc}, the penalty-based approach stands out for its greater flexibility, ease of implementation, and, in certain cases, better accuracy, which is also demonstrated solidly in \citet{spin-penalty-3}.

Accurate assessment of spin states and excited states can help us understand and predict various chemical and physical properties. 
For instance, in photochemistry and transition-metal complexes, reaction kinetics and dynamics are governed by crossings of potential energy surfaces from different spin states and excited states. 
Understanding the reaction mechanisms in photochemistry facilitates the design of novel light-harvesting and light-emitting materials, such as chromophores, perovskite solar cells, and organic light-emitting diodes (OLED)~\cite{Curutchet2016}.
In transition metal complexes, electronic spin can be manifested by unpaired electrons at metal centers or by redox non-innocent ligands. Highly accurate electronic structure methods empowers us to delicately control over their magnetism and catalytic reactivity~\cite{Swart2016}.
These potential advancements are expected to drive the development of new technologies and deepen our understanding across multiple scientific disciplines.

\section{Methods}
\subsection{LapNet ansatz}
In this work, we implement and test penalty methods on LapNet \cite{lapnet}, a state-of-the-art NNVMC ansatz in modeling the electronic structure of molecular systems.
LapNet utilizes the general attention mechanism\cite{transformer}, a module widely applied in the cutting-edge deep learning application such as GPT\cite{gpt}, to address the electron-electron correlation. It has demonstrated promising results across a wide range of chemical systems while keeping a feasible computational cost.

LapNet is a multi-determinant Slater-Jastrow-backflow type ansatz:
\begin{widetext}
    \begin{equation}\label{eq:ansatz}
        \Psi_{\theta}(\boldsymbol{X}) = e^{\mathcal{J}_\theta(\boldsymbol{X})} \sum_k^{N_{\det}} \det
        \begin{bmatrix}
            \phi^{k\uparrow}_{\theta, 1}
            (\boldsymbol{x}_1|\boldsymbol{x}_{\uparrow}, \boldsymbol{x}_{\downarrow}) & \cdots & \phi^{k\uparrow}_{\theta, 1}(\boldsymbol{x}_{N_{\uparrow}}|\boldsymbol{x}_{\uparrow}, \boldsymbol{x}_{\downarrow}) & \phi^{k\downarrow}_{\theta, 1}(\boldsymbol{x}_{N_{\uparrow} + 1}|\boldsymbol{x}_{\uparrow}, \boldsymbol{x}_{\downarrow}) & \cdots & \phi^{k\downarrow}_{\theta, 1}(\boldsymbol{x}_{N}|\boldsymbol{x}_{\uparrow}, \boldsymbol{x}_{\downarrow})\\
            \vdots & \ddots & \vdots & \vdots & \ddots & \vdots\\
            \phi^{k\uparrow}_{\theta, N}(\boldsymbol{x}_1|\boldsymbol{x}_{\uparrow}, \boldsymbol{x}_{\downarrow}) & \cdots & \phi^{k\uparrow}_{\theta, N}(\boldsymbol{x}_{N_{\uparrow}}|\boldsymbol{x}_{\uparrow}, \boldsymbol{x}_{\downarrow}) & \phi^{k\downarrow}_{\theta, N}(\boldsymbol{x}_{N_{\uparrow} + 1}|\boldsymbol{x}_{\uparrow}, \boldsymbol{x}_{\downarrow}) & \cdots & \phi^{k\downarrow}_{\theta, N}(\boldsymbol{x}_{N}|\boldsymbol{x}_{\uparrow}, \boldsymbol{x}_{\downarrow})
        \end{bmatrix},
    \end{equation}
\end{widetext}
where the input $\boldsymbol{X} = [\boldsymbol{x}_1, \cdots, \boldsymbol{x}_N]$ is composed of the spatial coordinates of $N$ electrons. The spatial coordinates should be partitioned into spin-up part $\boldsymbol{x}_{\uparrow}\equiv [\boldsymbol{x}_1, \cdots, \boldsymbol{x}_{N_{\uparrow}}]$ and spin-down part $\boldsymbol{x}_{\uparrow}\equiv [\boldsymbol{x}_{N_{\uparrow}+1}, \cdots, \boldsymbol{x}_{N}]$. These input features are fed into a permutation equivariant neural network $\boldsymbol{\phi}_{\theta}$, which consists of stacked Sparse Derivative Attention (SDA) blocks. SDA is a modified attention block that introduces more sparsity into the derivative, facilitating efficient Laplacian calculation.
The output of the equivariant component is divided into $N_{\det}$ square matrices, each containing $N$ rows and $N$ columns. 
Different from single-electron orbitals in the standard Slater determinant, each entry in the square matrix aggregates all electron information through the SDA block. This mechanism, known as backflow transformation \cite{backflow, backflow-1, backflow-2, backflow-3}, provides a powerful representation of the electron-electron correlation. 
The determinants of these matrices are summed and multiplied by the exponential Jastrow factor $\mathcal{J}_{\theta}(\bdX)$, yielding the value of the wave function.

\subsection{Loss function}
The loss function in our method is formulated as:
\begin{equation*}
   \mathcal{L}_{\theta} = E_{\theta}  + \omega_{\Splus}P_{\Splus} + \omega_{o}P_{\text{overlap}},
\end{equation*}
where $E_{\theta}$ denotes the energy expectation value, 
$E_{\theta} = \mathbb{E}_{\bdX \sim \Psi_{\theta}^2}[E_L(\bdX)]$; 
$P_{\Splus}$ is the $\Splus$ penalty enforcing spin symmetry;
$P_{\text{overlap}}$ represents the overlap penalty enforcing orthogonality between the training state and anchor states.  $\omega_{o}$ and $\omega_{\Splus}$ are weights used to balance different penalty terms. We use the KFAC optimizer \cite{kfac-alg} to minimize this loss function.

\subsection{Spin-related Penalties}\label{subsec:method_spin_penalty}
In most of real-space VMC studies, the ansatz $\Psi_{\theta}:\mathbb{R}^{3N}\rightarrow \mathbb{R}$ only represents a single component of the complete state. This component can generate the correct expectation of spin-independent operator. However, spin-dependent operators are not well-defined within this component. To handle spin-dependent operators, we can reconstruct the complete state by the antisymmetric property:
\begin{equation*}
   |\Psi_{\theta}\rangle = \frac{1}{C}\sum_{i\in\text{Sym}(N)} (-1)^{p_i} \hat{\mathbb{P}}_i [\Psi_{\theta}(\cdot)\ket{N_{\uparrow},N_{\downarrow}}],
\end{equation*}
where $|\Psi_{\theta} \rangle$ is the antisymmetric state and $C$ is the normalizing factor. 
$\text{Sym}(N)$ is the symmetric group with $N$ elements. $\ket{N_{\uparrow},N_{\downarrow}}=\ket{\uparrow}_1...\ket{\uparrow}_{N_{\uparrow}}\ket{\downarrow}_{N_{\uparrow}+1}...\ket{\downarrow}_N$  refers to the spin state where the first $N_{\uparrow}$ electrons are spin-up electron and the last $N_{\downarrow}$ electrons are spin-down electron. $\hat{\mathbb{P}}_i$ represents the permutation operators that reorder both the spatial coordinates and the associated spins orientation. $p_i$ is the parity of the permutation. 
The complete state $\ket{\Psi_{\theta}}$ is a common eigenstate of all the permutation operators $\hat{\mathbb{P}}_i$. Thus, as long as the operator $\hat{O}$ is commuted with permutation operators, the expectation of $\hat{O}$ can be computed through:
\begin{equation}
    \bra{\Psi_{\theta}}\hat{O}\ket{\Psi_{\theta}}=\sum_{n_{a}=0}^{N} C_N^{n_a}\braket{\Psi_{\theta}}{n_a, N-n_a}\bra{n_a,N-n_a}\hat{O}\ket{\Psi_{\theta}},\label{eq:sz_seperation}
\end{equation}
where $C_N^{n_a}=\frac{N!}{n_a!(N-n_a)!}$. Note that $\ket{\Psi_{\theta}}$ is also an eigenstate of $\hat{S}_z$ with eigenvalue of $\frac{N_{\uparrow}-N_{\downarrow}}{2}$, so only the $n_a=N_{\uparrow}$ term is non-zero. Applying this formula to $\hat{O}=\Ssquare$ will lead to:
\begin{equation*}
 \langle \Ssquare\rangle=\mathbb{E}_{\bdX\sim\Psi_{\theta}^2}S_{L}^2(\bdX),
\end{equation*}
where $S_L(\bdX)$ is defined in Equation \ref{eq:local-spin-square}. 
As discussed in the Section \ref{subsec:Splus}, the computational cost of this term is $\mathcal{O}(N)$ times more than the standard local energy evaluation.

Instead of directly computing the expectation of $\Ssquare$, we try to leverage the relation between $\Ssquare$ and $\Splus$ to accelerate the calculation:
\begin{equation*}
    \Ssquare = \hat{S}_z^2 + \hat{S}_z + \hat{S}_- \hat{S}_+
\end{equation*}
Note that $\hat{S}_-^{\dagger}=\Splus$, the expectation of $\Ssquare$ can be rewritten as:
\begin{equation*}
\begin{aligned}
\bra{\Psi_{\theta}}\Ssquare\ket{\Psi_{\theta}}=&\frac{(N_{\uparrow}-{N_{\downarrow}})(N_{\uparrow}-N_{\downarrow}+2)}{4} + \bra{\Psi_{\theta}}\hat{S}_- \hat{S}_+\ket{\Psi_{\theta}}\\
=&\frac{(N_{\uparrow}-{N_{\downarrow}})(N_{\uparrow}-N_{\downarrow}+2)}{4} + \braket{\hat{S}_+\Psi_{\theta}}{\hat{S}_+\Psi_{\theta}}.
\end{aligned}
\end{equation*}
When the last term goes to the zero, the expectation of $\Ssquare$ reaches the minimal value of $\frac{(N_{\uparrow}-{N_{\downarrow}})(N_{\uparrow}-N_{\downarrow}+2)}{4}$, which implies $\ket{\Psi_{\theta}}$ is an eigenstate of $\Ssquare$. This property makes $\braket{\hat{S}_+\Psi_{\theta}}{\hat{S}_+\Psi_{\theta}}$ a proper penalty term to enforce the spin symmetry. To numericaly compute the last term, we use the similar trick in Equation \eqref{eq:sz_seperation}. Remark that $\Splus$ increase the number of spin-up electron, so only the $n_{a}=N_{\uparrow}+1$ term is non-zero:
\begin{equation}
\label{eq:low_scale_square}
    \begin{aligned}
    \penaltya=&\braket{\hat{S}_+\Psi_{\theta}}{\hat{S}_+\Psi_{\theta}}\\
        =
    &C_N^{N_{\uparrow}+1}\braket{\hat{S}_+\Psi_{\theta}}{N_{\uparrow}+1,N_{\downarrow}-1}\braket{N_{\uparrow}+1,N_{\downarrow}-1}{\hat{S}_+\Psi_{\theta}}\\
    =&\mathbb{E}_{\bdX\sim \Psi_{\theta}^2}[R_{N_{\uparrow}+1}(\bdX)^2],
    \end{aligned}
\end{equation}
where $R_{N_{\uparrow}+1}(\bdX)$ is defined in Equation \eqref{eq:splus_res}. 

Although introducing this penalty term can lead to a solution with the correct spin symmetry, the training process can be unstable due to the large variance. To reduce the variance, we propose an improved $\Splus$ penalty:
\begin{equation}\label{eq:P+}
\begin{aligned}
        \penaltyb=(\mathcal{P}_{+})^2, \quad\mathcal{P}_{+}\triangleq\mathbb{E}_{\bdX\sim \Psi_{\theta}^2}[R_{\beta}(\bdX)],
\end{aligned}
\end{equation}
where $\beta$ can take any value from spin-down indices $N_{\uparrow}+1, N_{\uparrow}+2,..., N$. It is obvious that $P_{\Splus}$ is zero if $\mathbb{E}_{\bdX\sim \Psi_{\theta}^2}[R_{N_{\uparrow}+1}(\bdX)^2]=0$. We then show that $\ket{\Psi_{\theta}}$ is an eigenstate of $\Ssquare$ when $\penaltyb$ is zero. Due to the symmetry for all the spin-down electrons, $\mathcal{P}_{+}$ is in fact irrelevant to $\beta$. So we have:
\begin{equation*}
\begin{aligned}
        \mathcal{P}_{+}=&\frac{1}{N_{\downarrow}}\sum_{\beta}\mathbb{E}_{\bdX\sim \Psi_{\theta}^2}[R_{\beta}(\bdX)]\\
        =&\frac{1}{N_{\downarrow}}\mathbb{E}_{\bdX\sim \Psi_{\theta}^2}[N_{\downarrow}-\sum_{\alpha, \beta} \frac{\hat{\mathbb{P}}_{[\alpha, \beta]}[\Psi](\boldsymbol{X})}{\Psi(\boldsymbol{X})}]\\
        =&\frac{1}{N_{\downarrow}}\mathbb{E}_{\bdX\sim \Psi_{\theta}^2}[S_L^2(\bdX)-\frac{(N_{\uparrow}-{N_{\downarrow}})(N_{\uparrow}-N_{\downarrow}+2)}{4}]\\
        =&\frac{1}{N_{\downarrow}}[\langle \Ssquare\rangle-\frac{({N_{\uparrow}-N_{\downarrow}})(N_{\uparrow}-N_{\downarrow}+2)}{4}]
\end{aligned}
\end{equation*}
Thus, as long as $\mathcal{P}_{+}$ is zero, the expectation of $\Ssquare$ also reaches the minimal value of $\frac{({N_{\uparrow}-N_{\downarrow}})(N_{\uparrow}-N_{\downarrow}+2)}{4}$, which means the state is the eigenstate of $\Ssquare$.
Therefore $(\mathcal{P}_{+})^2$  is a valid penalty term to enforce spin symmetry.

Note that the empirical estimation of  $\mathcal{P}_{+}$ in Equation \eqref{eq:P+},
\begin{equation*}
 \tilde{\mathcal{P}}_{+}^{\beta}=\frac{1}{M}\sum_{\bdX_i\sim\Psi_{\theta}^2}R_{\beta}(\bdX_i)   ,
\end{equation*}
depends on the choice of $\beta$, even though $\mathcal{P}_{+}$ itself does not.
 In practice, to improve the numerical stability, we select the $\beta$ value from the spin-down electron indices in a round-robin manner during training.

\subsection{Overlap penalty}
 To obtain excited states within the variational Monte Carlo framework, several methods have been developed~\cite{excited_qmc_review}. These methods include state-averaged energy minimization \cite{state_average_umrigar, penalty_wagner1, penalty_wagner2, penalty_double, penalty_paulinet, penalty_NQS, aw} and state-specific variance minimization methods \cite{state_specific1, state_specific2, state_specific3}. Among them, the overlap penalty method \cite{penalty_wagner1, penalty_wagner2, penalty_paulinet} has been proven effective in targeting the lowest excited states of molecules. This kind of method tries to minimize both the energy of wave function and its overlap with lower energy states, thereby avoiding state collapse. 

In this study, to obtain the excited states within the same spin symmetry, we adopt the overlap penalty proposed by \citet{penalty_wagner1}, wherein the square of the pairwise overlap is directly employed to avoid the state collapse. The overlap penalty targeting the $i$th excited state can be expressed as:
\begin{equation*}
    P_{\text{overlap}} = \sum_{j<i} w_{ij}|S_{ij}|^2
\end{equation*}
where $S_{ij}$ denotes the normalized overlap between $\Psi_i$ and $\Psi_j$, $w_{ij}$ is a hyperparameter which should be larger than $E_i - E_j$ to avoid state collapse\cite{penalty_wagner1}. The square of overlap $|S_{ij}|^2$ can be estimated through Monte Carlo method using the following effective estimator \cite{estimator1, estimator2, penalty_paulinet}:
\begin{equation*}
\begin{aligned}
    |S_{ij}|^2 &= \frac{\langle\Psi_i|\Psi_j\rangle^2}{\langle\Psi_i|\Psi_i\rangle\langle\Psi_j|\Psi_j\rangle} \\&
    = \mathbb{E}_{\bdX\sim\Psi_j^2} \Big[\frac{\Psi_i(\bdX)}{\Psi_j(\bdX)}\Big]\mathbb{E}_{\bdX\sim\Psi_i^2} \Big[\frac{\Psi_j(\bdX)}{\Psi_i(\bdX)}\Big].
\end{aligned}
\end{equation*}
When the overlap penalty part is trained to be zero, the training state $\Psi_i$ is forced to be orthogonal with the anchor states $\Psi_j, (j < i)$.

\subsection{Gradient of penalty terms}
In this section, we present the form of penalty gradients in the loss function. 
The gradient for overlap penalty to target the $i$th excited state can be evaluated directly using Monte Carlo method:
\begin{equation*}
\begin{aligned}
    \frac{\partial P_{\text{overlap}}}{\partial \theta} &=\sum_{j < i}w_{ij}(V_{ij}    \mathbb{E}_{\bdX\sim\Psi_i^2}[(\frac{\Psi_j(\bdX)}{\Psi_i(\bdX)}-V_{ji})\partial_{\theta}\log\Psi_i(\bdX)]\\
    &\quad\quad +V_{ji}    \mathbb{E}_{\bdX\sim\Psi_j^2}[(\frac{\Psi_i(\bdX)}{\Psi_j(\bdX)}-V_{ij})\partial_{\theta}\log\Psi_j(\bdX)])
\end{aligned}
\end{equation*}
where $V_{ij}=\mathbb{E}_{\bdX\sim{\Psi_j^2}}[\frac{\Psi_i(\bdX)}{\Psi_j(\bdX)}]$, $\partial_{\theta} \log\Psi_i(\bdX)$ denotes the derivative of $\log\Psi_i(\bdX)$ with respect to network parameters $\theta$.

Due to the non-Hermitian property of raising operator $\Splus$, the gradient of our $\Splus$ penalty contains the gradient of local estimator. More concretely, the gradient of the $\Splus$ penalty can be calculated through the following rule:
\begin{equation*}
\begin{aligned}%
    &\partial_{\theta}\penaltyb = \partial_{\theta}(\mathcal{P}^{\beta}_{+})^2 = \partial_{\theta} \mathbb{E}_{\bdX\sim \Psi_{\theta}^2}[R_{\beta}(\bdX)]^2\\
    =&2\mathcal{P}^{\beta}_{+}\mathbb{E}_{\bdX\sim \Psi_{\theta}^2}[2(R_{\beta}(\bdX)-\mathcal{P}^{\beta}_{+})\partial_{\theta}\log\Psi_{\theta}(\bdX)+\partial_{\theta}R_{\beta}(\bdX)].
    \end{aligned}
\end{equation*}

\section*{Acknowledgements}
\begingroup
\footnotesize 
We want to thank ByteDance Research Group for inspiration and encouragement. This work is directed and supported by Hang Li and ByteDance Research. Ji Chen is supported by the National Key R\&D Program of China (2021YFA1400500) and National Science Foundation of China (12334003). Liwei Wang is supported by National Key R\&D Program of China (2022ZD0114900) and National Science Foundation of China (NSFC62276005).
\endgroup

\section*{Author contributions}
\begingroup
\footnotesize
Z.L., Z.X.L, R.L, L.W., J.C. and W.R. conceived the study; 
Z.L, Z.X.L and R.L. proposed algorithms;
Z.L, Z.X.L. and X.L. performed implementations;
Z.L, Z.X.L, R.L. and X.W. carried out simulations and analysis; 
Z.L, L.W., J.C. and W.R. supervised the project. 
Z.L., Z.X.L, R.L, X.L., X.W., L.W., J.C. and W.R. wrote the paper.
\endgroup

\normalem
\bibliography{ref}
\end{document}